\documentclass[12pt,preprint]{aastex}
\usepackage{bbm}
\usepackage{mathrsfs}
\usepackage{amssymb,amsmath,float}
\usepackage{rotating}
\usepackage{color}

\newcommand\de{\delta}
\newcommand\ep{\epsilon}

\newcommand\lam{\lambda}

\newcommand\ta{\tau}

\newcommand\om{\omega}

\renewcommand\>{\rangle}

\newcommand\ie{\emph{i.e.}}
\newcommand\eg{\emph{e.g.}}

\newcommand\beq{\begin{equation}}
\newcommand\eeq{\end{equation}}
\newcommand\bea{\begin{eqnarray}}
\newcommand\eea{\end{eqnarray}}
\newcommand\bal{\begin{align}}
\newcommand\eal{\end{align}}

\newcommand\fr{\frac}

\newcommand\ap{\approx}

\renewcommand\bal{\mbox{\boldmath$\alpha$}}

\begin{document}

\title{Exclusion of standard $\hbar\omega$ gravitons by LIGO observation}

\author{Richard Lieu$^1$}

\affil{$^1$Department of Physics and Astronomy, University of Alabama,
Huntsville, AL 35899\\}

\begin{abstract}
Dyson (2013) argued that the extraordinarily large number of gravitons in a gravitational wave makes them impossible to be resolved as individual particles.  While true, it is shown in this paper that a LIGO interferometric detector also undergoes frequent and {\it discrete} quantum interactions with an incident gravitational wave, in such a way as to allow the exchange of energy and momentum between the wave and the detector.  This opens the door to another way of finding gravitons.  The most basic form of an interaction is the first order Fermi acceleration (deceleration) of a laser photon as it is reflected by a test mass mirror oscillating in the gravitational wave, resulting in a frequency blueshift (redshift) of the photon depending on whether the mirror is advancing towards (receding from) the photon before the reflection.  If \eg~a blueshift occurred, wave energy is absorbed and the oscillation will be damped.  It is suggested that such energy exchanging interactions are responsible for the {\it observed} radiation reaction noise of LIGO (although the more common way of calculating the same amplitude for this noise is based on momentum considerations).  Most importantly, in each interaction the detector absorbs or emits wave energy in amounts far smaller than the standard graviton energy $\hbar\om$ where $\om$ is the angular frequency of the gravitational wave.  This sets a very tight upper limit on the quantization of the wave energy, {\it viz.} it must be at least $\ap 10^{11}$ times below $\hbar\om$, independently of the value of $\om$ itself.
\end{abstract}

\section{Introduction}

Since the landmark discovery of the first gravitational wave (GW) source, a binary black hole (BBH) merger, \cite{lig16a}, Advanced LIGO detected two more BBH mergers during its first observing run, \cite{lig16b, lig16c}, and another two during its second, \cite{lig17a,lig17b}.  Moreover, in this second run a binary neutron star (BNS) merger, GW170817, was discovered, \cite{lig17c}, with accompanying emission in various parts of the electromagnetic spectrum.  In particular, the 2~s delay w.r.t. the GW signal in the arrival time of the prompt high energy gamma-rays detected by GBM/Fermi, \cite{poz17}, puts an upper limit to any systematic deviation of the vacuum speed of GWs from light by 5 parts in $10^{16}$, because the distance to GW170104 is 42.5 Mpc from the optical counterpart observation, \cite{cou17} and \cite{del17}.  Yet another resembling feature between GW and light is the absence of dispersive propagation of the GW, \cite{lig17a}.  The result suggests that gravitons, like photons, are massless if they exist, \cite{lig16d}.

Indeed, with the specific properties of GW so revealed by the sources found to date, especially GW170817, it seems appropriate to ask if GW, like light, also comprises discrete units of energy, {\it viz.} gravitons.  At first instance, it is natural to conjecture that each graviton has the energy $\hbar\om$, similar to the photon, where $\om$ is the GW frequency.  Despite the persisting absence of a renormalizable theory of quantum gravity, \eg~\cite{sch14} and \cite{rod12}, quantizing gravitational waves in a flat background metric, with ensuing gravitons of energy $\hbar\om$ each, has been done a long time ago and is relatively simple and free from problems, \cite{gup52}.  In any case gravitons are a reasonable consequence of any serious quantum gravity theory.

The purpose of this paper is to supplement the general (and generally pessimistic) discussion of graviton detection prospects of \cite{rot06} and \cite{dys13} to demonstrate that, in addition to discovering GWs, Advanced LIGO observations can already be used to exclude the existence of $\hbar\om$ gravitons.  Our current approach differs from the previous ones in that while the latter relied upon the extremely and unobservably small Poisson fluctuation in the enormous number ($\ap 10^{37}$, \cite{dys13}) of GW gravitons LIGO expects to detect, this work takes into account the possibility of quantum {\it interactions} between the GW and the LIGO interferometer, interactions that involve real exchange of energy and momentum.  The idea to be presented is exceedingly simple, and does not necessitate any quantum field operators to model the GW.  Instead, we show that it is sufficient to adopt a purely {\it kinematic} argument, {\it viz.} because a freely suspending test mass has a continuum of energy levels (\cite{bra03}), if it is {\it observed} to be absorbing energy from, and re-emitting energy to, an incident GW in units far less than $\hbar\om$, there will be direct and irrefutable evidence in hand against at least this most naive scheme of GW quantization.

\section{Energy-momentum conserving interactions between a gravitational wave and an interferometric detector}

We develop our analysis with the specific design of LIGO in mind.  Consider two orthogonal laser beams of length $L$ each, and aligned with the $x$- and $y$-axes. Moreover, let the GW be a plane wave propagating along the $z$-axis.  Slowly moving masses in the Transverse-Traceless (TT) gauge obey the geodesic deviation equation, which simplifies to \beq \fr{d^2 \xi^k}{dt^2} = \fr{1}{2}\fr{d^2 h^{TT}_{jk}}{dt^2} \xi^j, \label{geodev} \eeq see \eg~\cite{gas13}.  For GWs having the $+$ polarization and wavelength $\lambda \gg L$, {\it viz.}~\beq h_{\mu\nu} = h_+ \ep_{\mu\nu} e^{-i\om t}, \label{hmn} \eeq small excursions $\xi$ about the unperturbed positions of the mirrors on the two axes satisfy the equations \beq \ddot\xi_x = \fr{1}{2} \om^2 hL e^{-i\om t};~\ddot\xi_y = -\fr{1}{2} \om^2 hL e^{-i\om t}, \label{xi} \eeq where $h=h_+$.

The energy of the test mass $M$ is \beq E = \fr{1}{2} M (|\dot\xi_x|^2 + |\dot\xi_y|^2) = \fr{1}{4} Mh^2 L^2\om^2, \label{E} \eeq
From (\ref{xi}), it is seen that the GW does not actually exchange energy with the test mass because the scalar product of the force and the velocity averages to zero; alternatively one can also appeal to the absence of a real part in $\ddot\xi_x \dot\xi_x$. Thus, it would appear that under the influence of a classical GW (or equivalently a GW comprising an irresolvably large number of gravitons) the mirror oscillates smoothly and without damping or growth; indeed it is shown in the next paragraph that the classical radiation reaction force also performs no net work, although that is no longer the case if quantum fluctuation in the force is taken into account.

Let us examine the situation of one laser photon reflected by the smooth surface of a test mass mirror moving at speed $v_x = \dot\xi_x$ towards the photon at the time of impact.  Assuming elastic scattering in the instantaneous frame of the test mass, it is readily shown (\cite{sau97}, \cite{ma15}) that to $O(v_x/c)$ the outgoing photon is blueshifted in the laboratory frame by the fractional amount $2v_x/c$, {\it viz.} from $\hbar\om_0$ to $\hbar\om_0 (1+2v_x/c)$.  The energy absorbed by the interferometric detector is ultimately drawn from the GW via the damping of the test mass's oscillation.  However, there is no time averaged energy gain or loss, since $\<v_x\> = 0$ in one cycle, corresponding to $\<N\>/2$ photons being blueshifted during the half cycle when the test mass was moving along the $-x$ direction and $\<N\>/2$ photons being redshifted by equal amounts during the next half cycle when the test mass was moving along $+x$.  Thus, while any instantaneous photon reflection event is accompanied by a real quantum interaction by which the interferometric detector exchanged an amount of energy \beq \ep = 2\hbar\om_0 \fr{v_x (t)}{c} \label{ep} \eeq with the GW,
after averaging over one cycle of GW oscillation there is no net energy loss or gain\footnote{To $O(v_x^2/c^2)$ there is systematic absorption of the GW energy that results in the damping of the test mass oscillation, but the effect is unobservably small being $\ap 10^{-41}$~W for LIGO, \cite{sau97}.} for the detector (same applies to the GW) to $O(v_x/c)$.  In fact, the effect in question is expressible in terms of the DC radiation reaction force doing work on the mirror, the rate of this process is the product of $2P/c$ (where $P$ is the power of the laser) and $v_x$, and vanishes when averaged over one GW oscillation cycle.

Yet when quantum mechanics is taken into account, fluctuations will ensure that the time averaged work done on the test mass (in the case of a sample average over finite time) does not exactly vanish, at least not for the radiation reaction force.  Even without enlisting the more sophisticated and somewhat artificial opto-mechanical damping dynamics of a detuned laser cavity of \cite{ma15}, the basic and inevitable effect of Fermi acceleration of the reflected photons by the test mass moving under the influence of an incident GW will ensure sufficiently frequent energy-momentum conserving interactions between the interferometer and the GW to seed fluctuation on all timescales.

Specifically, for a monochromatic GW with the above properties and spanning the duration $\ta$, the variance $(\de E_\ta)^2$ in the energy exchanged between the interferometer and the GW is \beq (\de E_\ta)^2 = \fr{4\hbar^2 \om_0^2}{c^2} \int_0^\ta v_x^2 (t) \lam dt = \fr{4\hbar^2 \om_0^2}{c^2} \lam\ta \<v_x^2\> = \fr{\hbar^2 \om_0^2}{c^2} N\om^2 h^2 L^2, \label{vartau} \eeq where $\lam$ is the mean photon arrival rate, $N=\lam\ta$, and $\<v_x^2\> = |\dot\xi_x|^2 = \om^2 h^2 L^2/4$ is obtained from (\ref{xi}).    As a result of this fluctuation, the standard deviation in the averaged test mass energy $E$ over one cycle of oscillation as given by (\ref{E}), is \beq \de E = \fr{2\pi}{\om\ta} \de E_\ta = \fr{2\pi \sqrt{N} \hbar \om_0 Lh}{c\ta}. \label{dE} \eeq  Dividing $\de E$ by $E$, one obtains \beq \fr{\de E}{E} = \fr{8\pi \sqrt{N} \hbar\om_0}{MLh\om^2 c\ta} = \fr{8\pi}{MLh\om^2 c} \sqrt{\fr{P\hbar\om_0}{\ta}}. \label{dEE} \eeq This leads in turn to a strain noise in $h$, via $\de E/E = \de h^2/h^2 = 2\de h/h$, of spectral amplitude at angular frequency $\om$ of \beq \tilde{h} = \sqrt{\ta}\de h =   \fr{4\pi\sqrt{P\hbar\om_0}}{ML\om^2 c}. \label{deh} \eeq The result differs from the \cite{wan13} expression of the amplitude of the strain noise due to radiation reaction by only a numerical factor of order unity.  Evidently the standard method of deriving this noise formula as presented in \eg~p16 of \cite{wan13}, which relies on momentum exchange between the GW and the interferometric detector, could also have enlisted the energy exchange consideration here.

As an estimate of the magnitude of the radiation reaction noise, we adopt the parameters 
of \cite{lig15} to arrive at $E \gg \hbar\om$ (\ie~the motion of the test mass is classical, so that the notion of a well defined $v_x (t)$ at any given $t$ is justified), and
\bea
\tilde h & = & 6.2 \times 10^{-24}~\left(\fr{P}{750~{\rm kW}}\right)^{1/2}
    \left(\fr{\lambda_0}{1064~{\rm nm}}\right)^{-1/2} \nonumber\\
    &&\quad \times \left(\fr{M}{40~{\rm kg}}\right)^{-1}
    \left(\fr{L}{4~{\rm km}}\right)^{-1}
    \left(\fr{\nu}{20~{\rm Hz}}\right)^{-2}~{\rm Hz}^{-1/2}
    \label{th}
\eea
When compared to the total noise curve in Figure 2 of \cite{lig15}, (\ref{th}) accounts for $\ap$~50 \% of the strain amplitude at $\nu=20$~Hz.  Thus radiation reaction noise is an important noise source in the range of frequencies relevant to LIGO (it can also be shown that the remaining 50 \% is due mostly to shot noise).  Note that this noise is the only direct and comprehensive evidence of the existence of the aforementioned interactions, because the $O(v_x/c)$ frequency shift of the photons is too small to have an interferometrically measurable impact.

\section{Limit on the existence of the standard graviton}

The key part of the previous section, from the viewpoint of testing the existence of standard gravitons in the GW, has to do with the energy exchange $\ep$ (see (\ref{ep})) between the interferometric detector and the GW every time a laser photon is reflected by the mirror, because these are genuinely quantum interactions in the sense that not only is energy emitted or absorbed by the detector in finite steps of $\ep$, each event is random and memoryless, being governed by the Poisson arrival times of the photons.

Yet the crucial deduction is made when one estimates $\ep$ for typical LIGO observing conditions, {\it viz.}~ \beq \ep = 3.12 \times 10^{-43}~\left(\fr{\lambda_0}{1064~{\rm nm}}\right)^{-1} \left(\fr{L}{4~{\rm km}}\right) \left(\fr{\nu}{20~{\rm Hz}}\right) \left(\fr{h}{10^{-21}}\right)~{\rm J}, \label{epJ} \eeq where use was made of the relation $\sqrt{\<v_x^2\>} = \om h L/2$ which is a consequence of (\ref{xi}).  Comparing (\ref{epJ}) to the energy $\hbar\om = 1.26 \times 10^{-32}$~J of the standard graviton at $\nu = 20$~Hz, we see that $\ep$ is smaller by the ratio \beq r = \fr{\ep}{\hbar\om} = 2.45 \times 10^{-11} ~\left(\fr{\lambda_0}{1064~{\rm nm}}\right)^{-1} \left(\fr{L}{4~{\rm km}}\right) \left(\fr{h}{10^{-21}}\right). \label{r} \eeq  Thus, the detector has been undergoing real quantum interactions with the GW in discrete amounts far less than the graviton, leading to the inference that for the GW to be quantized at all, the discreteness must occur in units $\ll \hbar\om$.

To be more explicit about the impossibility of standard gravitons in the wake of LIGO data, we emphasize that not only is (\ref{ep}) a consequence of basic conservation laws during the reflection of a photon, such energy-momentum exchanging interactions are the responsible cause of the radiation reaction noise in LIGO, {\it viz.} (\ref{th}), which is {\it observed}.  To reconcile theory with experiment, therefore, it is necessary that each graviton possesses an amount of energy $\lesssim \ep$, which means the GW quantization must occur at a level $\ap 10^{11}$ times smaller than the standard graviton.  It is also obvious from (\ref{r}) that our conclusion is independent of the frequency $\nu$ of the GW, \ie~so long as radiation reaction remains a significant noise component (which is always the case for LIGO's frequency range of GW visibility) this very stringent upper limit on the energy of the graviton applies.  Lastly, more complicated mechanisms of graviton absorption and emission, such as non-linear coupling between the photon and the graviton, will not be addressed in this paper, which focuses only on the most straightforward mode of detection similar to the discovery of the photon by the photoelectric effect.

The author thanks R.D. Blandford, Yuri Levin, Yiqiu Ma, and J.J. Quenby for helpful discussions.

\end{document}